\documentclass[floatfix,superscriptaddress,a4paper,
               nofootinbib,reprint]{revtex4}
\pdfoutput=1
\usepackage{graphicx}

\usepackage{hyperref}
\usepackage[latin2]{inputenc}
\usepackage{indentfirst}
\usepackage{enumerate}

\usepackage{amsmath}
\usepackage{amssymb}
\usepackage{url}

\newcommand{\eq}[1]{\begin{align} #1 \end{align}}

\newcommand{\ov}{\overline}

\begin{document}

\title{
Chemical freeze-out conditions in hadron resonance gas}

\author{V. Vovchenko}
\affiliation{
Frankfurt Institute for Advanced Studies, Goethe Universit\"at Frankfurt,
D-60438 Frankfurt am Main, Germany}
\affiliation{
Department of Physics, Taras Shevchenko National University of Kiev, 03022 Kiev, Ukraine}

\author{M.~I.~Gorenstein}
\affiliation{
Frankfurt Institute for Advanced Studies, Goethe Universit\"at Frankfurt,
D-60438 Frankfurt am Main, Germany}
\affiliation{Bogolyubov Institute for Theoretical Physics, 03680 Kiev, Ukraine}
\author{L.~M.~ Satarov}
\affiliation{
Frankfurt Institute for Advanced Studies, Goethe Universit\"at Frankfurt,
D-60438 Frankfurt am Main, Germany}
\affiliation{National Research Center ''Kurchatov Institute'', 123182 Moscow, Russia}

\author{H. Stoecker}
\affiliation{
Frankfurt Institute for Advanced Studies, Goethe Universit\"at Frankfurt,
D-60438 Frankfurt am Main, Germany}
\affiliation{
Institut f\"ur Theoretische Physik,
Goethe Universit\"at Frankfurt, D-60438 Frankfurt am Main, Germany}
\affiliation{
GSI Helmholtzzentrum f\"ur Schwerionenforschung GmbH, D-64291 Darmstadt, Germany}


\begin{abstract}
The hadron resonance gas  model with
the hadron-type dependent eigenvolume  corrections is employed to fit the hadron
yield data of the NA49 collaboration for central Pb+Pb collisions
at the center of mass energy of the nucleon pair
$\sqrt{s_{_{\rm NN}}} = 6.3, 7.6, 8.8, 12.3,$ and $17.3$~GeV, the hadron midrapidity yield
data of the STAR collaboration for Au+Au collisions at $\sqrt{s_{\rm NN}} = 200$~GeV,
and the hadron midrapidity yield data of the ALICE collaboration for Pb+Pb collisions at $\sqrt{s_{\rm NN}} = 2760$~GeV. The influence of the eigenvolume corrections is studied.
\end{abstract}

\maketitle

\section{Introduction}
Phenomenological thermodynamic models are very useful in extracting basic parameters
of the strongly interacting matter created in the relativistic
nucleus-nucleus (A+A) collisions,
particularly, to estimate its temperature~\cite{SOG1981, Molitoris1985,Hahn1987}.
The thermal parameters at chemical freeze-out -- the stage of A+A
collision when inelastic reactions between hadrons cease -- have been successfully
extracted by fitting the rich data on hadron yields in various experiments,
ranging from the low energies at SchwerIonen-Synchrotron (SIS) to the highest energy of
the Large Hadron Collider
(LHC), within the hadron resonance gas (HRG) model~\cite{CleymansSatz,CleymansRedlich1998,CleymansRedlich1999,
Becattini2001,Becattini2004,ABS2006,ABS2009}.
It has been argued~\cite{DMB}, that the
inclusion into the model of all known resonances as free non-interacting (point-like) particles allows to
effectively model the attraction between hadrons.
Such formulation, a multi-component point-particle gas of all known hadrons and resonances,
is presently the most commonly used one in the thermal model analysis.

In a realistic HRG model one also needs to take into account the
repulsive interactions between hadrons.
The HRG with the repulsive interactions have been successfully compared
with the lattice QCD data~\cite{EV-latt-1,EV-latt-2,EV-latt-3,VS2016}, and
it has recently been shown in Ref.~\cite{VS2015} that the inclusion of the
repulsive interactions into HRG in the form of a multi-component eigenvolume
procedure can significantly change the chemical freeze-out temperature while
improving the agreement with the ALICE hadron yield data compared to the point-particle HRG.
In the present work we perform a similar analysis at the finite (baryo)chemical
potential by considering the data on hadron yields in Pb+Pb and Au+Au collisions of NA49 and STAR collaborations.
In order to study the sensitivity of the obtained results
we use two different formulations of the multi-component eigenvolume HRG.

\section{Hadron Resonance Gas}
The  {\it ideal} HRG (I-HGR)
model corresponds to a statistical system of noninteracting
hadrons and resonances and leads to the following formula
for the system pressure in the grand canonical ensemble (GCE)
\begin{eqnarray}\label{id-press}
 P_{\rm I}(T,\mu_B)~&=&~\sum_j\,p_j^{\rm id}(T,\mu_j)~\nonumber
 \\
 &=&~
%
%
\sum_j \frac{d_j}{6\pi^2}\int_0^{\infty}\frac{k^4dk}{\sqrt{k^2+m_j^2}}
 \left\{\exp
\left[ \left( \sqrt{k^2+m_j^2}-\mu_j\right)/T  \right]\pm \eta_j\right\}^{-1}~,
\end{eqnarray}
where $d_j$ and $m_j$ are, respectively, the degeneracy factor and mass of $j$th
particle species, $\eta_j=-1$ corresponds to bosons and $\eta_j=1$ for ferimions
($\eta_j=0$ corresponds to the classical Boltzmann approximation).
The sum in Eq.~(\ref{id-press}) runs over all known hadron and resonances. The chemical potentials
$\mu_j$ for $j$th particle species are taken as
%
$ \mu_j~=~b_j\mu_B~+~s_j\mu_S~+~q_j\mu_Q~$,
%
where $b_j$, $s_j$, and $q_j$ correspond, respectively,
to the baryonic number, strangeness, and electric charge of $j$th particle;
chemical potentials $\mu_B,$, $\mu_S$, and $\mu_Q)$
regulate the average values of the conserved charges:  baryonic number $B$, strangeness $S$,
and electric charge $Q$.

In application to A+A collisions the free model parameters
of the HRG model are  $T$,
$\mu_B$, and $V$. The strange chemical
potential $\mu_S=\mu_S(T,\mu_B)$ and electric chemical
potential $\mu_Q=\mu_Q(T,\mu_B)$
are found from the conditions of zero net strangeness and fixed
proton-neutron ratio in the colliding nuclei, e.g., $Q/B \cong 0.4$
for heavy nuclei.
The other intensive thermodynamical functions,
like particle number densities,
are calculated from the pressure function by standard thermodynamic formulae.
Extensive quantities, like total numbers of particles, are obtained
by multiplying the corresponding densities by the system volume $V$.

The repulsive interactions between hadrons can be modeled by the eigenvolume correction
of the van der Waals type,
first proposed in Refs.~\cite{HagedornEV1,Gorenstein1981,Kapusta1983}, while
the thermodynamically consistent procedure for a single-component gas was formulated in Ref.~\cite{Rischke1991}.
In our study we use two different
formulations considered within
the Boltzmann statistics.
We expect that the effects of quantum statistics have a~minor
influence on the obtained results.

The single-component eigenvolume model of Ref.~\cite{Rischke1991} was generalized
to the multi-component case in Ref.~\cite{Yen1997}. It was assumed that the available volume
for each of the hadron species is the same, and equals to the total
volume minus sum of eigenvolumes of all the hadrons in the system.
This  leads to the trascendental equation
\begin{equation}
 \label{p}
 P(T,\mu_B)=\sum_j\,p_j^{\rm id}\left(T,\mu^*_j\right)~,~~~~~
 \mu^*_j=\mu_j -v_jP(T,\mu_B)~,
\end{equation}
with
$v_j=16\pi\,r_j^3/3$
being the eigenvolume parameter for the particle $j$.
At $v_j=0$
the EV-HRG model (\ref{p}) is reduced to the I-HRG model (\ref{id-press}).


Let us assume that we have $f$ different hadron species.
The pressure as function of the temperature and hadron densities has the following form
\eq{\label{eq:Pex1n}
P(T,n_1,\ldots, n_f) = T \sum_i \frac{n_i}{1 - \sum_j v_j n_j},
}
where the sum goes over all hadrons and resonances included in the model,
and where $v_i$ is the
eigenvolume parameter of hadron species $i$.
The eigenvolume parameter $v_i$ can be identified with
the 2nd virial coefficient of the single-component gas of hard spheres and
is connected to the hard-core hadron radius $r_i$ as
$v_i = 4 \cdot 4 \pi r_i^3 / 3$.
In the GCE one has to solve the non-linear equation
(\ref{p})
for the pressure.
The number densities in the GCE can be calculated as
\eq{\label{eq:nex1}
n_i(T, \mu_B) = \frac{n_i^{\rm id} (T, \mu_i^*)}{1 + \sum_j v_j n_j^{\rm id} (T, \mu_j^*)}.
}

The multi-component eigenvolume HRG model given by Eqs.~\eqref{eq:Pex1n}-\eqref{eq:nex1} is
the most commonly used one in the thermal model analysis.
Since this model does not consider the cross-terms in the virial expansion of the
multi-component gas of hard spheres (see details below) we will refer to it as the ``diagonal'' model.

The virial expansion of the classical (Boltzmann) multi-component gas of hard spheres up to 2nd order can be written as~\cite{LL}
\eq{\label{eq:virial}
P(T,n_1,\ldots, n_f) \cong T \sum_i n_i + T \sum_{ij} b_{ij} n_i n_j~,~~~~~
b_{ij} = \frac{2 \pi}{3} \, (r_i + r_j)^3
}
are the components of the symmetric matrix of the 2nd virial coefficients.

Comparing Eqs.~\eqref{eq:Pex1n} and \eqref{eq:virial} one can see that the diagonal
model is not consistent with the virial expansion of the multi-component gas
of hard spheres up to 2nd order and corresponds to a different matrix of 2nd virial coefficients, namely $b_{ij} = v_i$.
For this reason we
additionally consider the van der Waals like multi-component
eigenvolume model from Ref.~\cite{GKK}, which is formulated
in the GCE assuming Boltzmann statistics, and which is consistent
with the 2nd order virial expansion in Eq.~\eqref{eq:virial}. The pressure in this model reads as
\eq{\label{eq:Pex2n}
P(T,n_1,\ldots, n_f) = \sum_i P_i = T \sum_i \frac{n_i}{1 - \sum_j \tilde{b}_{ji} n_j},~~~~
\tilde{b}_{ij} = \frac{2\,b_{ii}\,b_{ij}}{b_{ii}+b_{jj}}
}
with $b_{ij}$ given by \eqref{eq:virial},
and where quantities $P_i$ can be regarded as ``partial'' pressures.
This eigenvolume model given by \eqref{eq:Pex2n} is initially formulated in the canonical ensemble.
In Ref.~\cite{GKK} it was transformed to the grand canonical ensemble.
In the GCE formulation one has to solve the following system of the non-linear
equations for $P_i$
\eq{\label{eq:Pex2pi}
P_i = p_i^{\rm id} (T, \mu_i - \sum_j \tilde{b}_{ij} \, P_j), \qquad i = 1, \ldots , f,
}
where $f$ is the total number of the hadronic components in the model.
The hadronic densities $n_i$ can then be recovered by solving the system
of linear equations connecting $n_i$ and $P_i$
\eq{\label{eq:Pex2ni}
T n_i + P_i \sum_j \tilde{b}_{ji} n_j = P_i, \qquad i = 1, \ldots , f~.
}
We refer to the model given by Eqs.~\eqref{eq:Pex2n}-\eqref{eq:Pex2ni} as
the ``crossterms'' eigenvolume model (see also
Ref.~\cite{BugaevEV}).
In practice, the
solution to \eqref{eq:Pex2pi} can be obtained by using an
appropriate iterative procedure. In our calculations
the Broyden's method~\cite{Broyden} is employed to
obtain the solution of the ``crossterms'' model, using
the corresponding solution of the ``diagonal'' model as the initial guess.

\section{Calculation results}
In our calculations we include strange and non-strange hadrons
listed in the Particle Data Tables~\cite{pdg}, along with their
decay branching ratios. This includes mesons up to $f_2(2340)$, (anti)baryons up to $N(2600)$.
We do not include hadrons with charm and bottom degrees of freedom which have a negligible
effect on the fit results, and we also removed the $\sigma$ meson ($f_0(500)$) and
the $\kappa$ meson ($K_0^*(800)$) from the particle list because
of the reasons explained in
Refs.~\cite{Broniowski:2015oha,Pelaez:2015qba}.
The finite width of the resonances is taken into account in the usual
way, by adding the additional integration over their Breit-Wigner shapes in the point-particle gas expressions.
The feed-down from decays of the unstable resonances to the total hadron yields is included in the standard way.

As was mentioned before, the inclusion of the eigenvolume interactions is one of the most popular extensions of the standard HRG model.
In most of the analyses dealing with chemical freeze-out which did include
the eigenvolume corrections~\cite{ABS2006,PHS1999,Cleymans2006} it was assumed that all the hadrons have the same eigenvolume.
It has been established that, in this case, the eigenvolume corrections can significantly reduce the densities~\cite{Begun2013,mf-2014}, and, thus, increase the total system volume at the freeze-out as compared to the point-particle gas at the same temperature and chemical potential.
For this parametrization, however, the eigenvolume corrections essentially cancel out in the ratios of yields and, thus, have a negligible effect on the values of the extracted chemical freeze-out temperature and chemical potential.
If, however, one considers hadrons with the different hard-core radii, then the ratios may change, and the fit quality can be improved~\cite{Gorenstein1999,BugaevEV}.

In order to test the sensitivity of the freeze-out conditions due to different hard-core hadron radii we consider two parametrizations.
Our main focus will be
on the~bag-model inspired parametrization, with the hadron eigenvolume proportional to its mass through a bag-like constant, i.e.,
\eq{\label{eq:BagEV}
v_i = m_i / \varepsilon_0\,.
}
Such eigenvolume parametrization had been obtained for the heavy Hagedorn resonances, and was used to describe their thermodynamics~\cite{HagedornEV1,Kapusta1983} as well as their
effect on particle yield ratios~\cite{NoronhaHostler2009}.
It was mentioned in the Ref.~\cite{Becattini2006} that such parametrization would lead to the increase of the freeze-out temperature, but that it does not entail an improvement of the fit quality in the ``diagonal'' EV model, or changes in other fit parameters.
Note that the eigenvolume for the resonances with the finite width is assumed to be constant for each resonance, and is determined by its pole mass.

We perform the thermal fit to the midrapidity yields of the charged pions,
charged kaons, (anti)protons, $\Xi^-$, $\Xi^+$, $\Omega$, $\bar{\Omega}$, $\Lambda$, $K^0_S$, and $\phi$,
measured by the ALICE collaboration in the 0-5\% most central Pb+Pb collisions
at $\sqrt{s_{\rm NN}} = 2.76$~TeV~\cite{ALICEdata}. Note that the
centrality binning for $\Xi$ and $\Omega$ hyperons is different
from the other hadrons. Thus, we take the midrapidity yields of $\Xi$
and $\Omega$ in the $0-5$\% centrality class from Ref.~\cite{Becattini2014},
where they were obtained using the interpolation procedure.

%
\begin{figure}[!t]
\centering
\includegraphics[width=0.49\textwidth,height=0.4\textwidth]{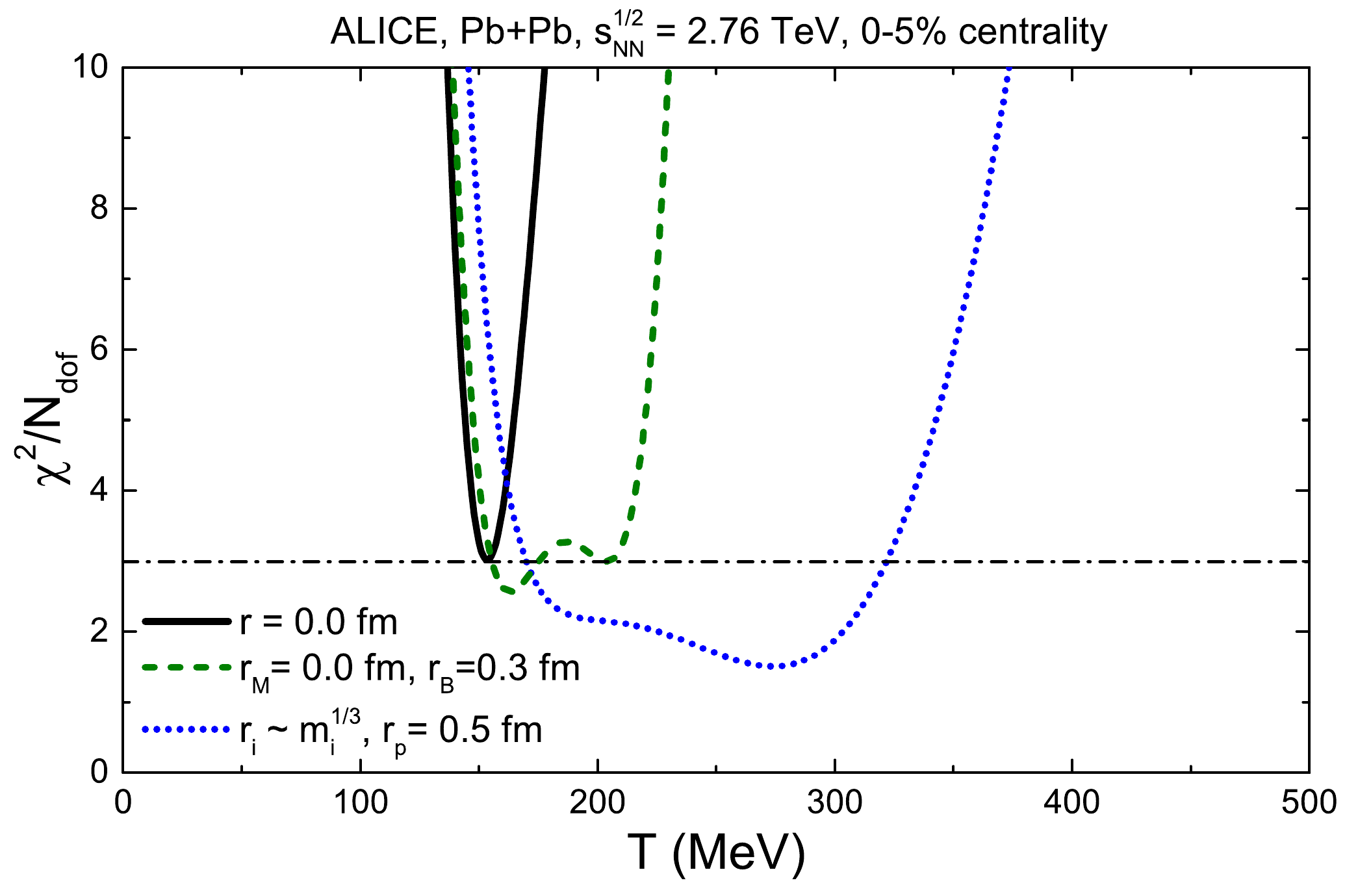}
\includegraphics[width=0.49\textwidth,height=0.38\textwidth]{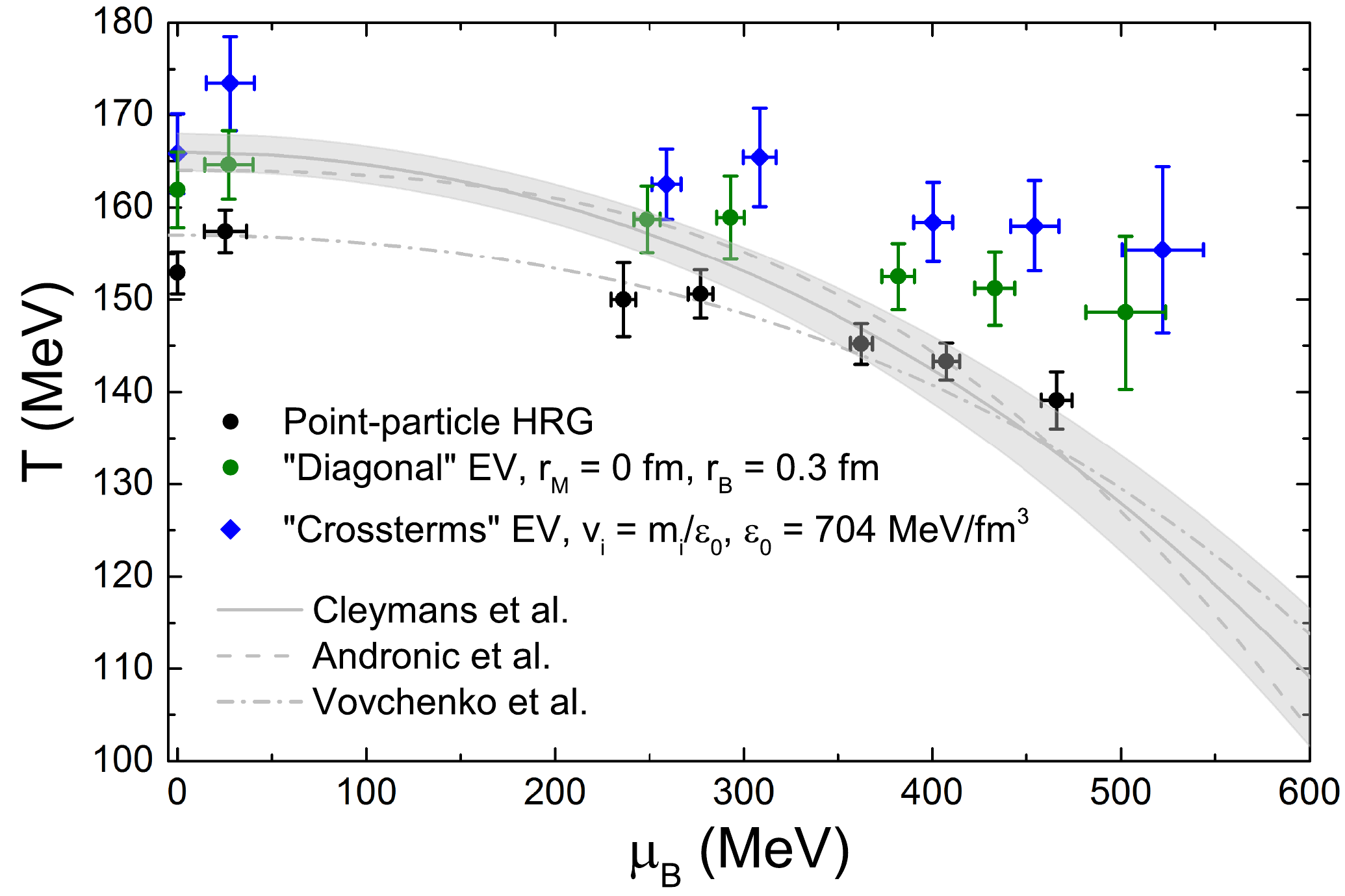}
\caption[]{
 Left: The temperature dependence of $\chi^2 / N_{\rm dof}$ of different fits to the ALICE data
on hadron yields in 0-5\% most central Pb+Pb collisions at 2.76~TeV. The solid line corresponds
to the point-particle HRG model, the dashed line to $r_B = 0.3$~fm and $r_M=0$,
and the dotted line to  bag-like eigenvolume HRG model
(the bag-like constant in Eq.~\eqref{eq:BagEV} is fixed to reproduce the hard-core proton radius of 0.5 fm.
Right: The freeze-out parameters within three different versions of the HRG:
point-particles, the ``diagonal'' EV model with $r_M=0$ and $r_B=0.3$~fm, and the
``crossterms'' eigenvolume HRG with $r_i \sim m_i^{1/3}$ and $r_p = 0.43$~fm.
The parameterized freeze-out curves from Refs.~\cite{ABS2009}, \cite{Cleymans2006}, and \cite{VBG2015},
obtained within the point-particle-like HRG models, are depicted by lines.
}
\label{fig:chi2-vs-T}
\end{figure}

The left Fig.~\ref{fig:chi2-vs-T} shows the temperature dependence of the $\chi^2/N_{\rm dof}$ for three
versions of the HRG model:
point-particle particles, i.e. all $v_i=0$,
the two-component eigenvolume HRG model with the point-like mesons $r_M=0$ and the
(anti)baryons of fixed size $r_B=0.3$ fm \cite{EV-latt-1},
and the bag-like eigenvolume HRG model
with the bag-like constant in Eq.~\eqref{eq:BagEV} fixed to reproduce the hard-core proton radius of 0.5~fm.
At each temperature the only remaining free parameter, namely the system volume
per unit slice of midrapidity,
is fixed to minimize the~$\chi^2$ at this temperature.

Presently not much is known about the eigenvolumes of different hadron species,
and there is no proof that parametrization \eqref{eq:BagEV} is the most realistic one.
For instance, it can be argued, that strange hadrons should have a different (smaller) eigenvolume compared to non-strange ones.
The bag-like constant $\varepsilon_0$ determines the magnitude of the hadron eigenvolumes.
The values of $r_p = 0.3-0.8$~fm have been rather
commonly used in the literature~\cite{Yen1997,BugaevEV,Begun2013,PHS1999,Cleymans2006,Gorenstein1999}.
Additionally, the value $r_p \simeq 0.6$~fm was extracted from the ground state properties of nuclear
matter within the fermionic van der Waals equation for nucleons~\cite{NM-VDW}.
Note that, within the bag-like parametrization, the hard-core radius $r_i$ of any hadron $i$
is related to the chosen value of $r_p$ through the relation $r_i = r_p \, \cdot \, (m_i/m_p)^{1/3}$,
where $m_i$ is the mass of the hadron $i$.


\begin{figure}[bht!]
\centerline{\includegraphics[trim=0 7.5cm 0 9cm, clip, width=0.8\textwidth]{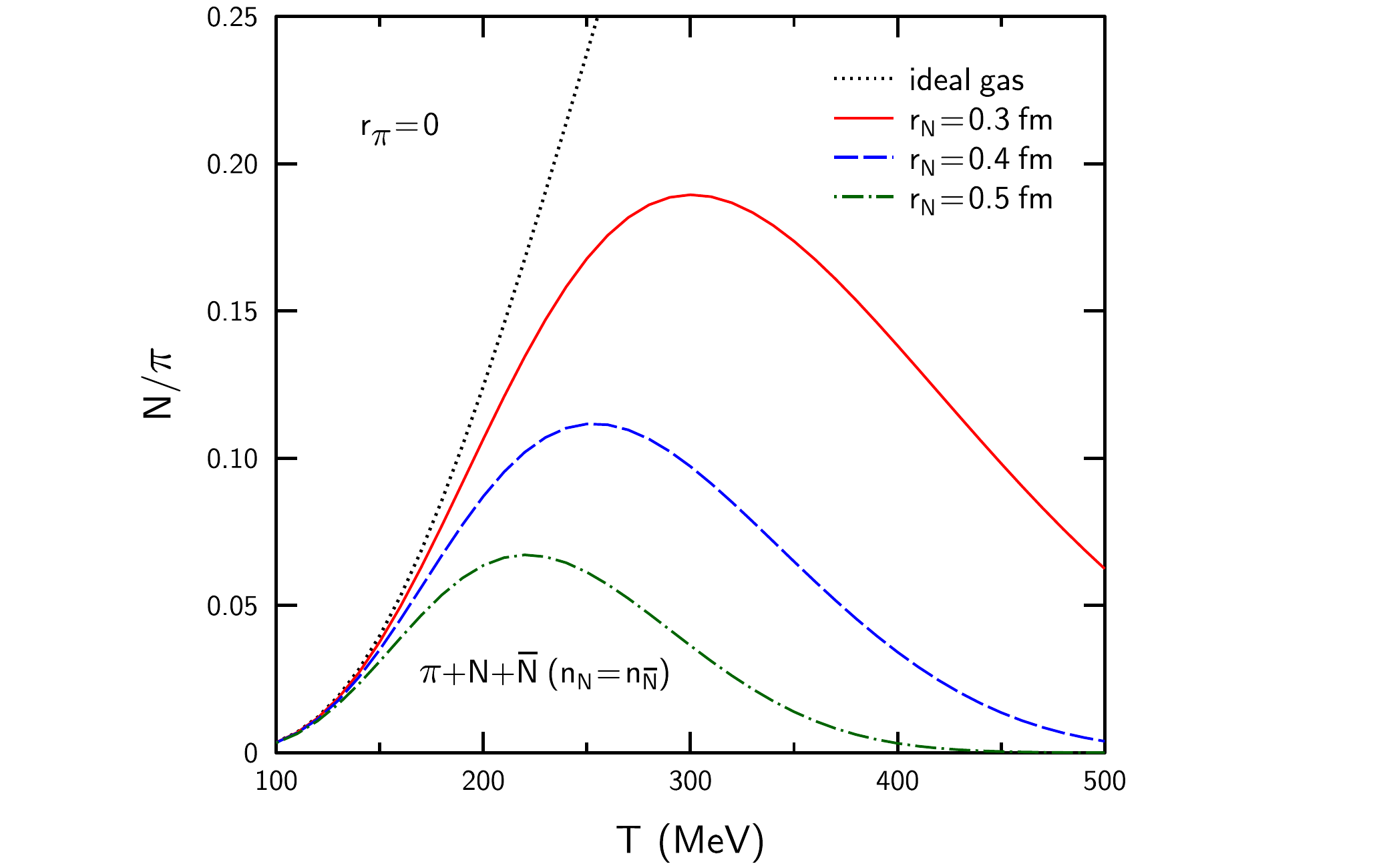}}
\caption[]{
The $N/\pi$ ratio as a function of temperature in the $\pi N\ov{N}$ matter with equal numbers
of nucleons and antinucleons. The thick solid, dashed, and dash-dotted lines correspond to
nucleon hard-core radii $r_N=0.3, 0.4$ and $0.5~\textrm{fm}$, respectively. The dotted line
is obtained in the ideal gas limiting case $r_N=0$.
}\label{fig:nrpi}
\end{figure}
As seen from the left Fig.~\ref{fig:chi2-vs-T} the $\chi^2/N_{\rm dof}$ has the second minimum
in the EV HRG models at high temperatures. To understand its physical origin let us consider
the EV HRG model with $r_M=0$ and $r_B=0.3$~fm for a simple case of the  $\pi N\ov{N}$ mixture with  with
$\mu_B=0$ (i.e. with equal average
numbers of nucleons and antinucleons.
Using the Boltzmann approximation, one obtains for the  nucleon to pion ratio:
\begin{equation}
\label{NNpi}
\frac{N_N(T)}{N_\pi(T)}~=~\exp\left[-~\frac{v_N \,P(T)}{T}\right]~\frac{n^{\rm id}_N(T)}{n^{\rm id}_\pi(T)}~,
\end{equation}
where pressure $P(T)$ is determined by solving Eq.~(\ref{p}) with $\mu_i=0\,(i=N,\ov{N},\pi)$ and $v_\pi=0$.
Figure~\ref{fig:nrpi} shows the temperature dependence of the $n_N/n_{\pi}$ ratio for several values of $r_N$.
One can see that the ratio (\ref{NNpi})  has a non-monotonic temperature dependence with a maximum
in the $T$-interval from 200 to 300~MeV.
An increase of $N_N/N_\pi$ at small $T$ is due to a strong increase
of $n^{\rm id}_N(T)\sim \exp(-m_N/T)$ at low temperatures. At large $T$ (and, thus, large system
pressure $P$)  the ratio (\ref{NNpi}) starts to decrease because of a stronger
EV suppression of (anti)nucleon densities as compared to pions.
At fixed value of $r_N$ one can fit the $N/\pi$
ratio by choosing two different values of temperature. The higher temperature value
corresponds to a denser state of the $\pi N\ov{N}$ system with strong short-range interactions of
mesons and (anti)baryons.


We perform the simultaneous fit of the hadron yield
data of the NA49, STAR, and ALICE collaborations.
The data of the NA49 collaboration includes
$4\pi$ yields of the charged pions, charged kaons, $\Xi^-$, $\Xi^+$, $\Lambda$, $\phi$, and, if
available, $\Omega$, $\bar{\Omega}$, measured in the 0-7\% most central Pb+Pb
collisions $\sqrt{s_{\rm NN}} = 6.3, 7.6, 8.8, 12.3$, and in the 0-5\% most central
Pb+Pb collisions at $\sqrt{s_{\rm NN}} = 17.3$~GeV~\cite{NA49data-1,NA49data-2}.
The feeddown from strong and electromagnetic decays is included in the model.
Additionally, the data on the total number of participants $N_W$ is identified with
total net baryon number and is included in the fit.
The actual tabulated data used in our analysis is available in Ref.~\cite{NA49data-3}.

The STAR data contains the midrapidity yields of charged pions, charged kaons,
(anti)protons, $\Xi^-$, $\Xi^+$, $\Omega$+$\bar{\Omega}$, and $\phi$ in the 0-5\% most central
Au+Au collisions at $\sqrt{s_{\rm NN}} = 200$~GeV~\cite{STARdata,BecattiniSTAR}.
The yield of protons also includes the feed-down from weak decays of (multi)strange hyperons,
this is properly taken into account in the model. We note that there is also available STAR
data on production of $\Lambda$ and $\bar{\Lambda}$.
These data are corrected for the feed-down from weak decays.
However, we have found that removing this feed-down in the model leads to a significant
worsening of the data description as compared to the case when weak decay feed-down
is included in the model. For this reason we decided to exclude yields of $\Lambda$ and $\bar{\Lambda}$ from the fit.
The eigenvolume HRG with $r_p = 0.4-0.6$~fm cannot satisfactorily
describe the lattice data at $T>200$~MeV and $\mu_B = 0$~MeV~\cite{EV-latt-3}.
Moreover, the high temperature $\chi^2$ minima shown in the left Fig.~\ref{fig:chi2-vs-T}
are plagued by several problems. Firstly, the speed of sound behaves unphysically,
namely, $c_s^2 \sim 1$ at the global minima for $r_p = 0.4-0.6$~fm.
The superluminal behavior of the speed of sound is a known problem of the EV model,
and avoiding it would require modifying the model.
Secondly, the packing fraction $\eta$ takes rather high values, typically
$\eta \sim 0.15$
at the best fit location.
At such high values of $\eta$ the eigenvolume model is expected to
deviate significantly from the equation of state of
the hard spheres model~(see, e.g., Refs.~\cite{mf-2014,Sat15,ZalewskiRedlich}).
To take care of the issues listed above
the additional constraint that the temperature range is restricted
to $T \lesssim T_0 \simeq 175$~MeV will be is now imposed.
One finds that the best fit
in such a scenario will approximately correspond to the first local minima shown
in the left Fig.~\ref{fig:chi2-vs-T} for $r_p = 0.40$~fm, with an increased values of
the chemical freeze-out temperature but with essentially unchanged $\chi^2$.

The inclusion of a bag-like eigenvolume leads to a better description
of the data at all the considered energies.
For the case when $r_M = 0$ and $r_B = 0.3$~fm the quality of description
of the data remains approximately the same compared to point-particle, with better
description at some energies, and worse at the others.
In both cases the inclusion of the finite eigenvolumes leads to some changes in the extracted parameters:
the chemical freeze-out temperature increases by about 10-15 MeV,
the baryochemical potential increases by about 10-15\% while the strangeness undersaturation parameter remains almost unaffected
(see the right Fig.~\ref{fig:chi2-vs-T}).
The fit errors of~$T$ and $\mu_B$,
obtained from analyses of the second-derivative error matrices at the minima,
increase notably for the finite EV cases.
The obtained results also indicate that the chemical
freeze-out curve in $T$-$\mu_B$ plane
has a smaller curvature in the EV models compared to the one
obtained within the point-particle HRG.
A similar result was obtained in \cite{Becattini2013} but by employing
a different mechanism, namely, by the considering the distortion
of yields due to the post-hadronization cascade phase.

The extraction of the chemical freeze-out parameters is thus rather sensitive
to the modeling of repulsive interactions between hadrons.
For these reasons,
even when the lattice constraint is used, the uncertainties in the extraction
of the chemical freeze-out parameters remain large.
These large uncertainties in the values of $T$ and $\mu_B$ at chemical freeze-out
seen in our analysis may indicate that the chemical freeze-out is not a sharp
process which takes place on some so-called freeze-out hypersurface with
very similar values of the temperature and chemical potential,
but that it is rather a continuous process, happening throughout the
whole space-time evolution of the system created in heavy-ion collisions,
and characterized by very different values of temperatures, energy densities,
and other parameters. Such a picture have been obtained within the transport model
simulations of the heavy-ion collisions by analyzing the space-time distribution
of the chemical ``freeze-out'' points of various hadrons~\cite{ContFrz}.

\section{Summary}

In summary, the data of the NA49, STAR, and ALICE collaborations on the hadron yields
in central Pb+Pb (Au+Au) collisions at $\sqrt{s_{\rm NN}} = 6.3, 7.6, 8.8, 12.3, 17.3$, $200$, and $2760$~GeV
is analyzed within the two different multi-component HRG models employing different eigenvolumes for different hadrons.
For the case of mass-proportional eigenvolumes,
fixed for a proton hard-core radius of 0.4-0.6~fm, these models describe the data
significantly better than the conventional point-particle HRG model in very wide regions in the $T$-$\mu_B$ plane.
Similarly, a much broader $\chi^2$ minima are observed when mesons are assumed
to be point-like while baryons have a fixed hard-core radius of $r_B = 0.3$~fm.

These results show that the extraction of the chemical freeze-out parameters
is extremely sensitive to the modeling of the short-range repulsion between the
hadrons, and imply that the point-particle HRG cannot be used for a reliable determination of the chemical freeze-out conditions.
Even within a more conservative approach, where we rather strictly
constrain the model parameters to the lattice data, we obtain a chemical
freeze-out curve which differs from the one obtained in the point-particle case,
has a systematically better fit quality of the data, and demonstrates a rather irregular
non-parabolic $\chi^2$ profile in the vicinity of the minima.
On the other hand,
the entropy per baryon extracted from the data for the different energies is found to be much more robust:
it is almost independent of the details of the modeling of the eigenvolume
interactions and of the specific $T-\mu_B$ values obtained.
This is consistent with the picture of continuous freeze-out, where hadrons are being frozen-out
throughout the extended regions of the space-time evolution of the system rather than from the sharp freeze-out hypersurface.

The obtained results demonstrate that inclusion of the
eigenvolume interactions are of crucial importance for thermal fitting the hadron yield data.
It is also shown that any conclusions based on thermal
fits should be based not just on the location of the $\chi^2$ minimum and its magnitude,
but rather on the full profile of the~$\chi^2$. In many cases the $\chi^2$ has an irregular
non-parabolic structure around the minimum, thus, the standard statistical-based estimates
of the uncertainties of the extracted parameters become inapplicable.

The collision energy range investigated in this work is relevant for the ongoing SPS and RHIC
beam energy scan programs, as well as for the experiments at the future FAIR and NICA facilities.
The presented results should be taken into account in the future analysis and interpretation of
the hadron yield data within these experiments.

\section*{Acknowledgements}
We dedicate this presentation to Walter Greiner. Scientific discussions
with him were really stimulating and guided our studies during last three decades.

\end{document}